\documentstyle[sprocl,epsf]{article}
\def\Journal#1#2#3#4{{#1} {\bf #2}, #3 (#4)}
\def\NPB{{\em Nucl. Phys.} B}
\def\PLB{{\em Phys. Lett.}  B}
\def\PRL{\em Phys. Rev. Lett.}
\def\PRD{{\em Phys. Rev.} D}
\def\ZPC{{\em Z. Phys.} C}
\def\ss{\scriptsize}

\def\be{\begin{equation}}
\def\ee{\end{equation}}
\def\bea{\begin{eqnarray}}
\def\eea{\end{eqnarray}}
\begin{document}
\begin{flushright}
TUM-HEP-291/97\\
July 1997\\
\end{flushright}
\vskip 1cm
\title{SUSY in Rare and CP-Violating $B$ Decays \footnote{Talk given by A.
Masiero at the 2nd International Conference on B Physics and CP
Violation (BCONF 97), Honolulu, HI, 24-28 Mar 1997.}}
\author{ A. MASIERO }
\address{SISSA -- ISAS, Trieste and Dip. di Fisica, Universit\`a di Perugia
and INFN, Sezione di Perugia, Via Pascoli, I-06100 Perugia, Italy}
\author{ L. SILVESTRINI }
\address{Physik Department, Technische Universit\"{a}t M\"{u}nchen, D-85748
Garching, Germany}
\maketitle\abstracts{
In this talk  we discuss rare $B$ decays ($b \to s
\gamma$, $b \to s g$, $b \to s \ell^{+} \ell^{-}$), $B-\bar{B}$ oscillations
and
CP violation in $B$ physics
in the context of low-energy SUSY. We outline the variety of predictions that
arise according to the choice of the SUSY extension ranging from what we call
the ``minimal" version of the MSSM to models without flavour universality or
with broken R-parity. In particular, we provide a model-independent
parameterization of the SUSY FCNC and CP-violating effects which is useful in
tackling the
problem in generic low-energy SUSY. We show how rare $B$ decays and CP
violation in $B$-decay amplitudes may be
complementary to direct SUSY searches at colliders, in particular for what
concerns extensions of the most restrictive version of the MSSM.}

\section{Introduction}
The generation of fermion masses and mixings (``flavour problem") gives
rise to a first and important distinction among theories of new physics
beyond the electroweak Standard Model (SM). Indeed, one may conceive a
kind of new physics which is completely ``flavour blind", i.e. new
interactions which have nothing to do with the flavour structure. To
provide an example of such a situation, consider a scheme where flavour
arises at a very large scale (for instance the Planck mass) while new
physics is represented by a supersymmetric (SUSY) extension of the SM
with supersymmetry broken at a much lower scale and with the SUSY
breaking transmitted to the observable sector by flavour blind gauge
interactions. In this case one may think that the new physics does not
cause any major change to the original flavour structure of the SM,
namely that the pattern of fermion masses and mixings is compatible with
the numerous and demanding tests of Flavour Changing Neutral Currents
(FCNC). Alternatively, one can conceive a new physics which is entangled
with the flavour problem. As an example consider a technicolour scheme
where fermion masses and mixings arise through the exchange of new gauge
bosons which mix together ordinary and technifermions. Here we expect
(correctly enough) new physics to have potential problems in
accommodating the usual fermion spectrum with the adequate suppression
of FCNC. As another example of new physics which is not flavour blind,
take a more conventional SUSY model which is derived from a
spontaneously broken N=1 supergravity and where the SUSY breaking
information is conveyed to the ordinary sector of the theory through
gravitational interactions. In this case we may expect that the scale at
which flavour arises and the scale of SUSY breaking are not so different
and possibly the mechanism itself of SUSY breaking and transmission is
flavour-dependent. Under these circumstances we may again expect
a potential flavour problem to arise, namely that SUSY contributions to
FCNC processes are too large.

The potentiality of probing SUSY in FCNC phenomena was readily realized when
the era of SUSY  phenomenology started in the early 80's.\cite{susy2}
 In particular, the
major implication that the scalar partners of quarks of the same electric
charge but belonging to different generations had to share a remarkably high
mass degeneracy was emphasized.

Throughout the large amount of work  in this last decade it became clearer
and clearer that generically talking of the implications of low-energy SUSY on
FCNC may be rather misleading. We have a minimal SUSY extension of the SM, the
so-called Minimal Supersymmetric Standard Model (MSSM),\cite{susy1}
where the FCNC
contributions can be computed in terms of a very limited set of unknown
new SUSY parameters. Remarkably enough, this minimal model succeeds to pass all
the set of FCNC tests unscathed. To be sure, it is possible to severely
constrain the SUSY parameter space, for instance using $b
\to s \gamma$ in a way which is complementary to what is achieved by direct
SUSY searches at colliders.

However,  the MSSM is by no means equivalent to low-energy SUSY. A first
sharp distinction concerns the mechanism of SUSY breaking and
transmission to the observable sector which is chosen. As we mentioned
 above, in models with gauge-mediated SUSY breaking it may
 be possible to avoid the
FCNC threat ``ab initio" (notice that this is not an automatic feature of
this class of models, but it depends on the specific choice of the
sector which transmits the SUSY breaking information, the so-called
messenger sector). The other more ``canonical" class of SUSY theories
that was mentioned above has gravitational messengers and a very large
scale at which SUSY breaking occurs. In this talk we will focus only on
this class of gravity-mediated SUSY breaking models. Even sticking to
this more limited choice we have a variety of options with very
different implications for the flavour problem.

\section{Rare $B$ decays in the MSSM and beyond}
\label{sec:rare}

Although the name seems to indicate a well-defined particle model,
actually
MSSM denotes at least two quite different classes of low-energy SUSY models.
In its most restrictive meaning it denotes the minimal SUSY extension of
the
SM (i.e. with the smallest needed number of superfields) with R-parity,
radiative breaking of electroweak symmetry, universality of the soft
breaking
terms and simplifying relations at the GUT scale among SUSY parameters. In
this ``minimal" version the MSSM exhibits only four free parameters in
addition to those of the SM. Moreover, some authors impose specific relations
between the two parameters $A$ and $B$ that appear in the trilinear and
bilinear scalar terms of the soft breaking sector, further reducing the number
of SUSY free parameters to three. Then, all SUSY masses are just function of
these few independent parameters and, hence, many relations among them
exist.
Obviously this very minimal version of the MSSM can be very predictive. The
most powerful constraint on this minimal model in the FCNC context comes from
$b \to s \gamma$.

In SUSY there are five classes of one-loop diagrams which contribute
to FCNC $B$
processes. They are distinguished according to the virtual particles running
in the loop: W and up-quarks, charged Higgs and up-quarks, charginos and
up-squarks, neutralinos and down-squarks, gluinos and down-squarks. It turns
out that, at least in this ``minimal" version of the MSSM, the charged Higgs
and chargino exchanges yield the dominant SUSY contributions. As for $b \to s
\gamma$ the situation can be summarized as follows. The CLEO
measurement~\cite{cleo}
yields BR$(B \to X_{s}\gamma)=(2.32 \pm 0.67)\times 10^{-4}$.
On the
theoretical side we have just witnessed a major breakthrough
with the
computation of the next-to-leading logarithmic result for the BR. This has
been
achieved thanks to the calculation of the $O(\alpha_{s})$ matrix
elements~\cite{greub} and
of the next-to-leading order Wilson coefficients at $\mu \simeq
m_{b}$.\cite{misiak} The
result quoted by Buras et al.~\cite{buras}
is BR$(B \to X_{s} \gamma)=(3.48 \pm 0.31)
\times 10^{-4}$ in the SM. 
A substantial improvement also on the experimental error is
foreseen for the near future. Hence $b \to s \gamma$ is going to constitute
the most relevant place in FCNC $B$ physics to constrain SUSY at least before
the advent of $B$ factories. So far this process has helped in ruling out
regions of the SUSY parameter space which are even larger than those excluded
by LEP I and it is certainly going to be complementary to what LEP II is
expected to do in probing the SUSY parameter space. After the detailed
analysis in 1991~\cite{bertol}
for small values of $\tan \beta$, there have been recent
analyses~\cite{barb}
  covering the entire range of $\tan \beta$ and including also other
technical improvements (for instance radiative corrections in the Higgs
potential). It has been shown~\cite{vissani}
that the exclusion plots are very sensitive also
to the relation one chooses between A and B. It should be kept in mind that
the ``traditional" relation $B=A-1$ holds true only in some simplified version
of the MSSM. A full discussion is beyond the scope of this talk and so we
refer
the interested readers to the vast literature which exists on the subject.

The constraint on the SUSY parameter space of the ``minimal" version of the
MSSM greatly affects also the potential departures of this model from the SM
expectation for $b \to s \ell^+ \ell^-$. The present
limits~\cite{cleo2}$^{\!-\,}$\cite{cdf} on the exclusive
channels BR$(B^{0} \to K^{*0} e^{+} e^{-})_{CLEO}<1.6 \times 10^{-5}$
 and BR$(
B^{0} \to K^{*0} \mu^{+} \mu^{-})_{CDF}<2.1 \times 10^{-5}$
are within an
order of magnitude of the SM predictions. On the theoretical side, it has
been estimated that the evaluation of $\Gamma (B\to X_{s}\ell^{+} \ell^{-})$ in
the
SM is going to be affected by an error which cannot be reduced to less than
$10-20 \%$ due to uncertainties in quark masses and interference effects from
excited charmonium states.\cite{ligeti}
It turns out that, keeping into account the bound
on $b \to s \gamma$, in the MSSM with universal soft breaking terms a $20 \%$
departure from the SM expected BR is kind of largest possible value one can
obtain.\cite{cho}
Hence the chances to observe a meaningful deviation in this case are
quite slim. However, it has been stressed that in view of the fact that three
Wilson
coefficients play a relevant role in the effective low-energy Hamiltonian
involved in $b \to s \gamma$ and $b \to s \ell^{+} \ell^{-}$, a third
observable in
addition to BR$(b \to s \gamma)$ and BR$(b \to s \ell^{+}\ell^{-})$ is needed.
This has been identified in some asymmetry of the emitted leptons (see
 refs.~\cite{cho}$^{\!-\,}$\cite{ali}
 for two different choices of such asymmetry).
This quantity, even
in the ``minimal" MSSM, may undergo a conspicuous deviation from its SM
expectation and, hence, hopes of some manifestation of SUSY, even in this
minimal realization, in $b \to s \ell^{+} \ell^{-}$ are still present.

Finally, also for the $B_{d}-\bar{B}_{d}$ mixing, in the above-mentioned
analysis of rare $B$ physics in the MSSM with universal soft breaking
terms~\cite{bertol}
it was emphasized that, at least in the low $\tan \beta$ regime, one cannot
expect an enhancement larger than $20\%-30\%$ over the SM prediction
(see also ref.~\cite{kurimoto}). Moreover
it was shown that $x_{s}/x_{d}$ is expected to be the same as in the SM.

It should be kept in mind that the above stringent results strictly depend not
only on the minimality of the model in terms of the superfields that are
introduced, but also on the ``boundary" conditions that are chosen.
All the low-energy SUSY masses are computed in terms of the $M_{Pl}$ four
SUSY parameters through the RGE evolution. If one relaxes this tight
constraint on the relation of the low-energy quantities and treats the masses
of the SUSY particles as independent parameters, then much more freedom is
gained. This holds true even if flavour universality is enforced. For
instance,
 BR$(b \to s \gamma
)$ and $\Delta m_{B_{d}}$ may vary a lot from the SM expectation, in
particular in regions of moderate SUSY masses.\cite{brignole}

Moreover, flavour universality is by no means a prediction of low-energy SUSY.
The absence of flavour universality of soft-breaking terms may result from
radiative effects at the GUT scale or from effective supergravities derived
from string theory. In the non-universal case,
 BR$(b \to s \ell^{+} \ell^{-})$
is strongly affected by this larger freedom in the parameter space. There are
points of this parameter space where the nonresonant BR$(B \to X_{s} e^{+}
e^{-})$ and BR$(B \to X_{s} \mu^{+}\mu^{-})$  are enhanced by up to $90 \%$ and
$110 \%$ while still respecting the constraint coming from $b \to s
\gamma$.\cite{cho}

\section{Model-independent analysis of FCNC processes in SUSY}

Given a specific SUSY model it is in principle possible to make a full
computation of all the FCNC phenomena in that context. However, given the
variety of options for low-energy SUSY which was mentioned in the Introduction
(even confining ourselves here to models with R matter parity), it is
important to have a way to extract from the whole host of FCNC processes a set
of upper limits on quantities which can be readily computed in any chosen SUSY
frame.

The best model-independent parameterization of FCNC effects is the so-called
mass insertion approximation.\cite{mins}
It concerns the most peculiar source of FCNC SUSY contributions that do not
arise from the mere supersymmetrization of the FCNC in the SM. They originate
from the FC couplings of gluinos and neutralinos to fermions and
sfermions.\cite{FCNC} One chooses a basis
for the fermion and sfermion states where all the couplings of these particles
to neutral gauginos are flavour diagonal, while the FC is exhibited by the
non-diagonality of the sfermion propagators. Denoting by $\Delta$ the
off-diagonal terms in the sfermion mass matrices (i.e. the mass terms relating
sfermions of the same electric charge, but different flavour), the sfermion
propagators can be expanded as a series in terms of $\delta = \Delta/
\tilde{m}^2$
where $\tilde{m}$ is the average sfermion mass.
As long as $\Delta$ is significantly smaller than $\tilde{m}^2$,
we can just take
the first term of this expansion and, then, the experimental information
concerning FCNC and CP violating phenomena translates into upper bounds on
these $\delta$'s.\cite{deltas}$^{\!-\,}$\cite{ggms}

  Obviously the above mass insertion method presents the major advantage that
 one does not need the full diagonalization of the sfermion mass matrices to
 perform a test of the SUSY model under consideration in the FCNC sector. It is
 enough to compute ratios of the off-diagonal over the diagonal entries of the
 sfermion mass matrices and compare the results with the general bounds on the
 $\delta$'s that we provide here from all available experimental information.

  There exist four different
$\Delta$ mass insertions connecting flavours $i$ and $j$
 along a sfermion propagator: $\left(\Delta_{ij}\right)_{LL}$,
$\left(\Delta_{ij}\right)_{RR}$, $\left(\Delta_{ij}\right)_{LR}$ and
$\left(\Delta_{ij}\right)_{RL}$. The indices $L$ and $R$ refer to the
helicity of
the
 fermion partners. The size of these $\Delta$'s can be quite different. For
 instance, it is well known that in the MSSM case, only the $LL$ mass insertion
 can change flavour, while all the other three above mass insertions are
flavour
 conserving, i.e. they have $i=j$. In this case to realize a $LR$ or $RL$
flavour
 change one needs a double mass
insertion with the flavour changed solely in a $LL$
 mass insertion and a subsequent flavour-conserving $LR$ mass insertion.
Even worse
 is the case of a FC $RR$ transition: in the MSSM this can be accomplished only
 through a laborious set of three mass insertions, two flavour-conserving $LR$
transitions and an $LL$ FC insertion.
  Instead of the dimensional quantities $\Delta$ it is more
useful to provide bounds making use of dimensionless quantities, $\delta$,
that are obtained dividing the mass insertions by an average sfermion mass.

The FCNC processes in $B$ physics which provide the best bounds on the
$\delta_{23}$ and $\delta_{13}$ FC insertions are $b \to s \gamma$ and $B_{d}-
\bar{B}_{d}$, respectively.

The process $b \to s \gamma$ requires a helicity flip. In the presence of a
$\left(\delta^d_{23}\right)_{LR}$ mass insertion we can realize this flip in
the gluino running in the loop. On the contrary, the $\left(
\delta^d_{23}\right)_{LL}$ insertion requires the helicity flip to occur in
the external $b$-quark line. Hence we expect a stronger bound on the
$\left(\delta^d_{23}\right)_{LR}$ quantity. Indeed, this is what happens:
$\left(\delta^d_{23}\right)_{LL}$ is essentially not bounded, while
$\left(\delta^d_{23}\right)_{LR}$ is limited to be $<(1.3 \div 3)
 \times 10^{-2}$
for an average squark mass of 500 GeV and $0.3 < m^2_{\tilde g}/m^2_{\tilde q}
<4.0$ (these bounds scale as $m^2_{\tilde q}$).
Given the upper
bound on $\left(\delta^d_{23}\right)_{LR}$ from $b \to s \gamma$, it turns out
that the quantity $x_{s}$ of $B_{s}-\bar{B}_{s}$ mixing receives
contributions from this kind of mass insertions which are very tiny. The only
chance to obtain large values of $x_s$ is if $\left(\delta^d_{23}\right)_{LL}$
is large, say of $O(1)$. In that case $x_s$ can easily jump up to values of $O
(10^{2})$ or even larger.

As for the mixing $B_{d}-\bar{B}_{d}$, we obtain
\bea
\sqrt{\left\vert {\mbox Re} \left(\delta^d_{13}\right)^{2}_{LL}\right\vert}
&<&4.6 \cdot 10^{-2}\, ; \nonumber \\
\sqrt{\left\vert {\mbox Re} \left(\delta^d_{13}\right)^{2}_{LR}\right\vert}
&<&5.6 \cdot 10^{-2}\, ; \nonumber \\
\sqrt{\left\vert {\mbox Re} \left(\delta^d_{13}\right)_{LL}\left(
\delta^d_{13}\right)_{RR}\right\vert}
&<&1.6 \cdot 10^{-2}\, ;
\label{limbdbdb}
\eea
for $x\equiv m^{2}_{\tilde{g}}/m^{2}_{\tilde{q}}=0.3$ with $m_{\tilde{q}}=500$
 GeV. The above bounds scale with $m_{\tilde{q}}$(GeV)$/500$ for different
values of $m_{\tilde{q}}$ (at fixed $x$).\\
Then, imposing the bounds~(\ref{limbdbdb}), we can obtain the largest possible
value for BR($b \to d \gamma$) through gluino exchange. As expected, the
$\left( \delta^{d}_{13}\right)_{LL}$ insertion leads to very small values of
this BR of $O(10^{-7})$ or so, whilst the $\left( \delta^{d}_{13}\right)_{LR}$
insertion allows for BR($b \to d \gamma$) ranging from few times $10^{-4}$ up
to few times $10^{-3}$ for decreasing values of $x=m^{2}_{\tilde{g}}/
m^{2}_{\tilde{q}}$. In the SM we expect~\cite{ali96}
BR($b \to d \gamma$) to be typically $10-20$
times smaller than BR($b \to s \gamma$), i.e. BR($b \to d \gamma)=(1.7\pm 0.85
)\times 10^{-5}$. Hence a large enhancement in the SUSY case is conceivable if
$\left( \delta^{d}_{13}\right)_{LR}$ is in the $10^{-2}$ range. Notice that in
the MSSM we expect $\left( \delta^{d}_{13}\right)_{LR}<m^{2}_{b}/
m^{2}_{\tilde{q}}\times V_{td}<10^{-6}$, hence with no hope at all of a
sizeable contribution to $b \to d \gamma$.

\section{Is it possible to disentangle SUSY from SM in CP-violating $B$
decays?}

Keeping on with the approach that tackles FCNC effects in a generic SUSY
extension of the SM, we would like to address the following two questions
in this section: i) how large are the uncertainties of the SM predictions
for CP asymmetries in B decays? and ii) in which processes and how can one
possibly distinguish SUSY from SM contributions (without making any
commitment to the particular SUSY model)? \cite{cptutti}

Concerning the former above question we will work in the framework of the
analysis of ref.~\cite{cpnoi}  We use the
effective Hamiltonian
(${\cal H}_{eff}$) formalism, including
LO QCD corrections;  in the numerical analysis, we use the LO SM Wilson
coefficients evaluated at $\mu=5$ GeV, as given in ref.~\cite{zeit}
In most  of the cases, by choosing different scales (within a resonable range)
or by using NLO Wilson coefficients,  the
results vary by about $20-30 \%$ .  This is true
with the  exception of some particular channels where uncertainties  are
larger. The matrix elements of the operators of ${\cal H}_{eff}$
are given in terms of the following Wick contractions
 between hadronic states: Disconnected
Emission ($DE$), Connected Emission ($CE$), Disconnected Annihilation ($DA$),
Connected Annihilation ($CA$), Disconnected Penguin ($DP$)
and Connected Penguin
($CP$) (either for left-left ($LL$) or for left-right ($LR$)
current-current operators). Following ref.~\cite{cfms}, where
a detailed discussion can be found,
instead of  adopting a specific model for  estimating  the different
diagrams,  we let them vary within reasonable
ranges to estimate the SM uncertainty (see ref.~\cite{cpnoi}).
First,
only $DE=DE_{LL}=DE_{LR}$ are assumed to be different from zero
(for simplicity, unless stated otherwise,  the same numerical
values are used for  diagrams corresponding to the insertion
of  $LL$ or  $LR$ operators, i.e. $DE=DE_{LL}=DE_{LR}$,
$CE=CE_{LL}=CE_{LR}$, etc.).  We then consider,
in addition to $DE$,  the $CE$ contribution  by taking
 $CE=DE/3$.
Annihilation diagrams  are then included,
and we use  $DA=0$ and $CA=1/2 DE$.\cite{cfms}
Inspired by kaon decays, we allow for some enhancement of
the matrix elements  of left-right (LR) operators and choose
$DE_{LR}=2 DE_{LL}$ and $CE_{LR}=2 CE_{LL}$.
Penguin contractions, $CP$ and $DP$,
can be interpreted  as  long-distance penguin contributions to the matrix
elements and play an important role:  if   we
take $CP_{LL}=CE$ and $DP_{LL}=DE$, in some decays
these terms  dominate the amplitude.
Finally,  we allow for long distance
effects which might differentiate penguin contractions with up and charm quarks
in the loop, giving rise to incomplete GIM cancellations (we assume
$\overline{DP}= DP(c) - DP(u) =  DE/3$ and
$\overline{CP}= CP(c) - CP(u) =CE/3$).
For any given decay channel, whenever two terms
with  different CP phases contribute in the SM, letting the various matrix
elements vary within the above ranges, we estimate the ratio $r_{SM}$ of
the two amplitudes, which is reported in the fifth column of
table~\ref{tab:results}.

New physics changes SM predictions on  CP asymmetries in $B$ decays in
two ways:
by shifting the phase of the $B_{d}$--$\bar{B}_{d}$ mixing  amplitude
and by modifying both  phases and absolute values of  the decay ones.
The generic SUSY extension of the SM considered here affects all these
quantities.

In the SUSY case, by using for the Wilson coefficients in eq.~(12)
the results of ref.~\cite{ggms}
and by parameterizing the matrix elements  as we did for the SM case
discussed above,  we obtain the ratios of  SUSY to SM
 amplitudes given in table~\ref{tab:results}. For each decay channel we
give results  for squark and gluino masses of 250 GeV ($r_{250} $
in the sixth column). We remark that the inclusion of the various terms
in the amplitudes, $DE$, $DA$, etc.,
can  modify the ratio $r$ of  SUSY to SM contributions up to one
order of magnitude.

In terms of the decay amplitude $A$, the CP asymmetry reads
\begin{equation}
{\cal A}(t) = \frac{(1-\vert \lambda\vert^2) \cos (\Delta M_d t )
-2 {\rm Im} \lambda \sin (\Delta M_d t )}{1+\vert \lambda\vert^2}
\label{eq:asy}
\end{equation}
with $\lambda=e^{-2i\phi^M}\bar{A}/A$.
In order to be able to discuss the results  model-independently,
we have labelled as $\phi^M$ the generic  mixing phase.
The ideal case occurs when  one  decay
amplitude only appears in (or dominates)
a decay process: the CP violating asymmetry is  then determined by the
 total phase   $\phi^T=\phi^M+\phi^D$, where $\phi^D$  is the weak phase
  of the decay.
This ideal situation is spoiled by the presence of
several interfering amplitudes.
If the ratios $r$ in table~\ref{tab:results} are small, then the uncertainty on
the sine of the CP phase is $<  r $, while if $r$ is O(1)  $\phi^T$ receives,
in general,  large corrections.
\par
The results of our analysis are summarized in table~\ref{tab:results}.
In the third column, we give the values for the BR's of the various
channels,\cite{bkpi}
obtained using QCD sum rules form factors~\cite{qcdsr} to evaluate $DE$, and
$\vert CE \vert = 0.46 \times DE$, 
fitted using the measured two-body $B$ decays. The
range of values given corresponds to the variation of the CKM angles and to the
inclusion of the charming penguin contributions (see ref.~\cite{bkpi} for
further details).
In the fourth column, for each channel, we give  the possible SM decay phases
when one or two decay amplitudes contribute, and the range of variation of
their
ratio,  $r_{SM}$,   as explained above.
A few comments are necessary at this point:
a) for  $B \to K_S \pi^{0}$ the
 penguin contributions (with a vanishing phase) dominate over the
tree-level amplitude   because the latter is Cabibbo suppressed;
b) for the channel $b
\to s \bar s d$  only penguin operators or penguin contractions of
current-current operators  contribute; c) the phase $\gamma$ is present in the
penguin contractions of the $(\bar b u)(\bar u d)$ operator,
denoted as $u$-P $\gamma$
in table~\ref{tab:results}; d)
 $\bar b d \to \bar q q $ indicates processes occurring via annihilation
 diagrams which can be measured
 from the last two channels of table~\ref{tab:results};
e) in the case $B \to K^{+} K^{-}$ both
current-current and penguin operators contribute; f) in $B \to D^{0} \bar
D^{0}$ the contributions
from  the $(\bar b u) (\bar u d)$ and the   $(\bar b c) (\bar c d)$
current-current operators   (proportional to the phase $\gamma$)
tend to cancel out.

SUSY contributes to the decay amplitudes with  phases
induced by  $\delta_{13}$ and
$\delta_{23}$ which we denote as $\phi_{13}$ and $\phi_{23}$. The ratio
$A_{SUSY}/A_{SM}$ for SUSY masses of 250 GeV
 is reported in the $r_{250}$ column
of table~\ref{tab:results}.

We now draw some conclusions from the results of table~\ref{tab:results}.
In the SM, the first
six  decays  measure directly the mixing phase $\beta$, up to
corrections which, in most of the cases, are expected to be small.
These corrections, due to the presence of  two
amplitudes contributing with different phases,  produce
 uncertainties of $\sim 10$\% in   $B \to K_S \pi^{0}$,
 and  of $\sim 30$\%  in $B \to D^{+} D^{-}$ and $B \to
J/\psi \pi^{0}$.   In spite
of the uncertainties,  however, there are cases where
 the SUSY contribution gives rise to significant changes.
 For example, for SUSY masses of O(250) GeV, SUSY corrections  can  shift the
measured value of the sine of the phase in
 $B \to \phi K_S$ and in $B \to K_S \pi^{0}$ decays by an amount of
 about 70\%.  For these decays  SUSY effects are sizeable even for
masses of 500 GeV.  In $B \to
J/\psi K_S$  and $B \to \phi \pi^0$ decays, SUSY effects are only  about $10$\%
but SM uncertainties are negligible.  In $B \to K^0 \bar{K}^0$
the larger  effect, $\sim 20$\%,   is partially covered by the
indetermination of
about $10$\%  already existing in the SM.
Moreover the rate for this channel is expected to be rather small.
In $B \to D^{+} D^{-}$  and $B \to K^{+} K^{-}$, SUSY effects are
completely obscured  by the errors in the estimates of the SM amplitudes.
In $B^0\to D^0_{CP}\pi^0$ the asymmetry  is sensitive to the mixing angle
$\phi_M$ only because the decay amplitude is unaffected by SUSY.
This result can be used in connection with $B^0 \to K_s \pi^0$, since
a difference in the measure of the phase  is  a manifestation
of SUSY effects.
\par
Turning to $B \to \pi \pi$ decays, both the uncertainties
in the SM  and  the SUSY contributions are very large. Here we
witness the presence of three independent amplitudes with different phases
and of comparable size.  The observation of SUSY effects in
the $\pi^{0} \pi^{0}$ case is hopeless. The possibility of
separating SM and SUSY contributions  by using the isospin
analysis remains an open possibility which deserves further investigation.
For a thorough discussion of the SM uncertainties in $B \to \pi \pi $ see
ref.~\cite{cfms}

In conclusion, our analysis shows that measurements of CP asymmetries in
several channels may allow the extraction of the CP mixing phase and
to disentangle  SM and SUSY contributions to the CP decay phase.
The golden-plated decays in this respect are $B \to \phi K_S$
and $B \to K_S \pi^0$ channels. The size of the SUSY effects is
clearly controlled by the the non-diagonal SUSY mass
insertions $\delta_{ij}$, which for illustration we have assumed to have the
maximal value compatible with the present experimental limits on
$B^0_d$--$\bar B^0_d$ mixing.

\begin{table}[t]
 \begin{center}
 \caption[]{Branching ratios and CP phases for B decays. $\phi^{D}_{SM}$
 denotes the decay phase in
 the SM; T and P denote Tree and Penguin, respectively; for each
 channel, when two amplitudes with different weak phases are present,
 one is given in the first row, the other in the last one
 and the ratio of the two  in the $r_{SM}$ column. $\phi^{D}_{SUSY}$
 denotes the phase of the SUSY amplitude, and the ratio of the SUSY to SM
 contributions is given in the $r_{250}$ column.}
 \label{tab:results}
 \begin{tabular}{|ccccccc|}
 \hline
 \ss{Incl. }&\ss{ Excl. }&\ss{ BR $\times 10^{5}$ }&\ss{ $\phi^{D}_{\rm SM}$ }&
 \ss{ $r_{\rm SM}$ }&\ss{
 $\phi^{D}_{\rm SUSY}$ }&\ss{ $r_{250}$ }\\ \hline
 \ss{ $b \to c \bar c s$ }&\ss{ $B \to J/\psi K_{S}$ }&\ss{ $40$}&\ss{ 0 }&
 \ss{ -- }&\ss{$\phi_{23}$ }&\ss{ $0.03-0.1$ }\\ \hline
 \ss{ $b \to s \bar s s$ }&\ss{ $B \to \phi K_{S}$ }&\ss{ $0.6-2$}&\ss{ 0 }&
 \ss{ -- }&\ss{
 $\phi_{23}$ }&\ss{ $0.4-0.7$ }\\ \hline
 \ss{$b \to u \bar u s$ }&\ss{ }&\ss{ }&\ss{ P $0$ }&\ss{  }&\ss{  }&\ss{  }\\
 \ss{
  }&\ss{$ B \to \pi^{0} K_{S}$ }&\ss{ $0.02 - 0.4$ }&\ss{  }&\ss{ $0.01-0.08$
}&
  \ss{$\phi_{23}$ }&\ss{ $0.4-0.7$ }\\
  \ss{
 $b \to d \bar d s$ }&\ss{ }&\ss{ }&\ss{ T $\gamma$ }&\ss{  }&\ss{  }&\ss{  }
 \\ \hline
 \ss{
 $b \to c \bar u d$ }&\ss{ }&\ss{ }&\ss{ 0 }&\ss{  }&\ss{  }&\ss{  }\\
 \ss{
  }&\ss{$ B \to D^{0}_{CP} \pi^{0}$ }&\ss{ $16$}&\ss{  }&\ss{ 0.02 }&\ss{ -- }&
  \ss{ -- }\\
  \ss{
 $b \to u \bar c d$ }&\ss{ }&\ss{ }&\ss{ $\gamma$ }&\ss{  }&\ss{  }&\ss{  }\\
 \hline
 \ss{
  }&\ss{ $B \to D^{+} D^{-}$ }&\ss{ $30-50$}&\ss{ T $0$ }&\ss{ $0.03-0.3$ }&
  \ss{  }&\ss{
  $0.007-0.02$ }\\
  \ss{
  $b \to c \bar c d$}&\ss{ }&\ss{ }&\ss{ }&\ss{  }&\ss{ $\phi_{13}$ }&\ss{ }\\
  \ss{
  }&\ss{ $B \to J/\psi \pi^{0}$ }&\ss{ $2$}&\ss{ P $\beta$ }&\ss{ $0.04-0.3$ }&
  \ss{
  }&\ss{ $0.007-0.03$ }\\ \hline
  \ss{
  }&\ss{ $B \to \phi \pi^{0}$ }&\ss{ $1-4 \times 10^{-4}$ }&\ss{ P $\beta$}&
  \ss{ -- }&\ss{ }&\ss{
  $0.06-0.1$ }\\
  \ss{
  $b \to s \bar s d$}&\ss{ }&\ss{  } & &\ss{ }&\ss{ $\phi_{13}$ }&\ss{ }\\
  \ss{
  }&\ss{ $B \to K^{0} \bar{K}^{0}$ }&\ss{ $0.007-0.3$}&\ss{ {\it u}-P
  $\gamma$
  }&\ss{ $0-0.07$ }&\ss{ }&\ss{ $0.08-0.2$ }\\ \hline
  \ss{
 $b \to u \bar u d$ }&\ss{ $B \to \pi^{+} \pi^{-}$ }&\ss{ $0.2-2$ }&\ss{ T
 $\gamma$
 }&\ss{ $0.09-0.9$ }&\ss{ $\phi_{13}$ }&\ss{ $0.02-0.8$ }\\
 \ss{
 $b \to d \bar d d$ }&\ss{ $B \to \pi^{0} \pi^{0}$ }&\ss{ $0.003-0.09$ }&\ss{
 P $\beta$ }&\ss{ $0.6-6$
 }&\ss{ $\phi_{13}$ }&\ss{ $0.06-0.4$ }\\ \hline
 \ss{
 }&\ss{ $B \to K^{+} K^{-}$ }&\ss{ $< 0.5 $ }&\ss{ T $\gamma$ }&\ss{ $0.2-0.4$
}&\ss{ }&\ss{
 $0.04-0.1$ }\\
 \ss{
 $b \bar d \to q \bar q$ }&\ss{ }&\ss{  }&\ss{ }&\ss{ }&\ss{ $\phi_{13}$}&\ss{
}\\
 \ss{
 }&\ss{ $B \to D^{0} \bar D^{0}$ }&\ss{ $<20$ }&\ss{ P $\beta$
 }&\ss{ only $\beta$ }&\ss{  }&\ss{ $0.01-0.03$ }\\  \hline
 \end{tabular}
 \end{center}
 \end{table}

\section*{Acknowledgements}

We thank  M. Ciuchini, E. Franco, F. Gabbiani, E. Gabrielli and G. Martinelli
who collaborated with us in the model-independent analyses that we
presented here.
We wish to thank the organizers of $B$ Physics and CP Violation Conference
 for giving us the opportunity of discussing $B$
physics in a  stimulating and pleasant atmosphere.
The work of A. M. was partially supported by the EU contract ERBFMRX CT96
0090. L. S. acknowledges the support of Fondazione Angelo della Riccia,
Firenze.

\end{document}